%% file: LGI_ambiguous_final.tex
\documentclass[twocolumn,pra,aps,a4paper,floatfix,10pt]{revtex4-1}
\usepackage{amsmath}
\usepackage{graphicx}
\usepackage{amssymb}
\usepackage{xspace}

\usepackage{bbm}

\usepackage{color}

\usepackage[T1]{fontenc}

\newcommand{\rb}[1]{\left( #1 \right)}
\newcommand{\ew}[1]{\left\langle #1 \right\rangle}
\newcommand{\beq}{\begin{eqnarray}}
\newcommand{\eeq}{\end{eqnarray}}
\newcommand{\op}[2]{| #1 \rangle \langle #2 |}
\newcommand{\eq}[1]{Eq.~(\ref{#1})}
\newcommand{\fig}[1]{Fig.~\ref{#1}}
\newcommand{\secref}[1]{Sec.\,\ref{#1}}
\newcommand{\trace}[1]{\mathrm{Tr}\left\{#1\right\}}

\newcommand{\citer}[1]{{Ref.~\cite{#1}}}
\newcommand{\A}[1]{{\bf (A#1)}\xspace}

\begin{document}

\title{Ambiguous measurements, signalling and violations of Leggett-Garg inequalities}

\author{Clive Emary}
\affiliation{
 Joint Quantum Centre (JQC) Durham-Newcastle, School of Mathematics, Statistics and Physics, Newcastle University, Newcastle-upon-Tyne, NE1 7RU, United Kingdom
  }

\begin{abstract}
Ambiguous measurements do not reveal complete information about the system under test.  Their quantum-mechanical counterparts are semi-weak (or in the limit, weak-) measurements and here we discuss their role in tests of the Leggett-Garg inequalities.
We show that, whilst ambiguous measurements allow one to forgo the usual non-invasive measureability assumption, to derive an LGI that may be violated, we are forced to introduce another assumption that equates the invasive influence of ambiguous and  
unambiguous detectors. 
Based on this assumption, we derive signalling conditions that should be fulfilled for the plausibility of the Leggett-Garg test.  We then propose an experiment on a three-level system with a direct quantum-optics realisation that satisfies all signalling constraints and violates a Leggett-Garg inequality.
\end{abstract}

\maketitle

\section{Introduction\label{SEC:intro}}

The Leggett-Garg inequalities (LGIs) \cite{Leggett1985} were constructed as tests of \textit{macrorealism}, as defined by the three assumptions (the first two as stated in \citer{Leggett1985}; the third was made explicit in e.g. Refs.~\cite{Leggett2008,Kofler2008,Kofler2013}):
\begin{enumerate}
  \item[\A{1}] \textit{Macroscopic realism per se}: A macroscopic system with two or more macroscopically distinct states available to it will at all times {\em be} in one or the other of these states;
  \item[\A{2}] \textit{Noninvasive measurability (NIM)} at the macroscopic level: It is possible, in principle, to determine the state of the system with arbitrarily small perturbation on its subsequent dynamics;
  \item[\A{3}] \textit{Arrow of time}: The outcome of a measurement on the system cannot be affected by what  will or will not be measured on it later.
\end{enumerate}
The inequalities that follow from these assumptions have been the subject of much work that was reviewed a few years ago in \citer{Emary2014} with many theoretical \cite{Asadian2014,Schild2015,Brierley2015,Halliwell2016,Halliwell2016a,Lambert2016,Clemente2016} and experimental \cite{Robens2015,Zhou2015a,Robens2016,Knee2016,Formaggio2016a,Huffman2016,Wang2017,Wang2017a,Katiyar2017}  developments having taken place since.

Of the three assumptions, (A2) is particularly vexatious since, while it assumes NIM to hold in principle, non-invasivity must also be seen to hold true in practice, otherwise  any violation of an LGI can be assigned to some unwitting invasivity of the measurement \cite{Montina2012}.  This is the ``clumsiness loophole'' of Wilde and Mizel \cite{Wilde2012}.  And, since quantum-mechanical measurements are in fact invasive, NIM is a counterfactual and can never be ruled out empirically.

Leggett and Garg's proposal \cite{Leggett1985} for dealing with this situation was to use ideal negative measurements, and these have been employed in various recent experiments \cite{Knee2012,Katiyar2013,Robens2015,Katiyar2017,Wang2017}.  This approach, however, just shifts the locus of any presumed non-invasivity away from the system itself and onto some degree of freedom in its environment.  There have also been a number of attempts to formulate LGIs under different assumptions \cite{Huelga1995,Foster1991,Elby1992,Lapiedra2006,Zukowski2010}, but these too must ultimately suffer from similar loopholes.

This issue of invasivity relates to an important difference between the LGIs and the formally-similar Bell's inequalities, and this is the issue of \textit{signalling} \cite{Kofler2013,Halliwell2016}.  For spacelike-separated observers Alice and Bob in a Bell test, we have the no-signalling condition
\beq
  P(A) - \sum_B P(A,B) = 0
  \label{EQ:NS}
  ,
\eeq
where $P(A)$ is the probability that Alice obtains result $A$ and $P(A,B)$ is the joint probability of result $A$ for Alice and $B$ for Bob.  Thus, the influence of Bob's measurement is statistically undetectable to Alice (and vice versa). 
In the LGI setting, there is no external physical principle such as locality to which we can appeal that enforces a lack of signalling between the two measurements.
Let us define
\beq
  \delta(n_3) = P(n_3) - \sum_{n_2} P(n_3,n_2)
  \label{EQ:firstdeltadefn}
  ,
\eeq
where $n_2$ and $n_3$ are outcomes of measurements at times $t_2$ and $t_3>t_2$, to quantify the signalling in a LGI context. Under the assumptions (A1--3), these signalling quantifiers should be zero, just as in the Bell's test. Indeed the \textit{no-signalling-in-time} (NSIT) equalities  $\delta(n_3) =0 $ \cite{Kofler2013}, or ones very similar to them \cite{Li2012}, have been discussed as tests of macrorealism themselves \cite{Kofler2013,Emary2014b,Schild2015,Clemente2016}.  These studies show that, generically, quantum-mechanical violations of an LGI are accompanied by a violation of NSIT conditions (see \citer{George2013} and also \secref{SEC:noNIM}).  
From this, a  macrorealist would conclude that, since there is experimental evidence that the measurements can signal forward in time, they are invasive and (A2) does not hold in practise.  
Thus, observation that the NSIT equalities hold may be taken as a necessary [but by no means  sufficient] condition that our measurements appear non-invasive.  NSIT also restores the symmetry between LGIs and Bell inequalities.  

A number of ways of achieving LGI violations without signalling have been discussed.
Much has been written about weak (or semi-weak) measurements \cite{Aharonov1988,Kofman2012} and the violation of the LGIs \cite{Jordan2006,Ruskov2006,Williams2008,Palacios-Laloy2010,Goggin2011,Dressel2011,Suzuki2012,Bednorz2012}.  As Halliwell makes clear \cite{Halliwell2016},  weakly-measured quasiprobalities naturally have the NSIT property and can violate LGI inequalities.  The importance of weak measurements in unifying spatial and temporal correlations, and hence LG and and Bell inequalities, has been discussed in e.g. \citer{Marcovitch2011,*Marcovitch2011a}. 
Outside the weak-measurement paradigm, the work of George et al. \cite{George2013} (see also \citer{Maroney2012}) stands out as having reported a measurement of an LGI violation whilst obeying the  relevant no-signalling conditions (George et al. used the language of the measurements being ``non-disturbing'' rather than no-signalling).  These results, however, came in a very specific setting, viz the quantum 3-box paradox, leaving open the question as to whether their results represent a peculiarity of this model, or whether similar situations can be found in other, perhaps even macroscopic, systems.

In this paper we describe a general approach to LGI violations using a set of ambiguous measurements \cite{Dressel2012} that are realised quantum-mechanically by a particular class of POVM.  We show that with the quasiprobabilities inferred from such measurements we are able to violate an LGI inequality without making the NIM assumption.  
However, in order to justify use of these quasiprobabities as proxies for the real thing, we find we must make an alternative assumption that equates the invasive influence of ambiguous and unambiguous detectors on a macroreal state.  We will call this assumption \textit{Equivalently-invasive measureability} (EIM).
We believe that an assumption along these lines tacitly underlies previous work on weak measurements and the LGI. 
While it may seem hard to justify a priori that two potentially very different detectors are invasive in the same way (although, see later), this assumption leads to a testable consequence, namely that the signalling quantifiers for both detectors should be equal.  These conditions we call \textit{equal-signalling in time} (ESIT) and they can be tested empirically.  We note here that the assumption of EIM in the presence of fulfilled NSIT conditions is similar to the non-collusion of adroit measurements discussed by Wilde and Mizel \cite{Wilde2012}.

From our considerations we arrive at several conclusions: 
(i)\,Violations of any single LGI are always accompanied by violations of NSIT (this is a slight generalisation of a result given in \cite{George2013});
(ii)\,There exists specific combinations of quantum dynamics and ambiguous measurements that can both violate LGIs and satisfy ESIT. 
In the examples we discuss here, ESIT is realised through the NSIT conditions for both ambiguous and unambiguous each being exactly zero.  We thus show that the results of George can be generalised to arbitrary systems, provided we choose the dynamics and measurements appropriately.
(iii)\,In general, violations of LGI in the weak-measurement limit are accompanied by a violation of ESIT (and would thus be unconvincing to a macrorealist).  This must always be the case when our measurements involve just two outcomes. However, we also show that there exist specific weak-measurement scenarios with multiple outcomes in which ESIT remains intact.

This paper proceeds as follows.
In \secref{SEC:noNIM} we derive a modified version of the LGI without making the NIM assumption and show that directly-measured probabilities, even those from quantum mechanics, can never violate it.
In \secref{SEC:ambig} we then introduce our ambiguous measurements,  discuss the necessity of the EIM assumption and construct our ambiguously-measured LGI.
We then analyse this quantum-mechanically in \secref{SEC:quantum} and show how the ambiguous LGI can be violated whilst ESIT is preserved in general. 
In \secref{SEC:invmeasure} we discuss a concrete example of our formalism, eminently-realisable in terms of the quantum optics set up pursued in Refs.\,\cite{Wang2017,Wang2017a}.  In \secref{SEC:weak} we make the connection with weak measurements, before discussing the significance of our results in \secref{SEC:disc}.

\section{The LGI without the non-invasive measurability assumption \label{SEC:noNIM}}

The most-studied LG correlator involves dichotomic observable $Q=\pm 1$ and reads 
\beq
  K \equiv \ew{Q_2 Q_1} + \ew{Q_3 Q_2} - \ew{Q_3 Q_1}
  ,
\eeq
where $Q_i = Q(t_i)$ is the measurement outcome at times $t_3>t_2>t_1$. Under assumptions A1-3 above, Leggett and Garg showed that $K\le1$.  We want to investigate this correlator without assumption (A2).
To simplify matters, we first assume the coincidence of the measurement at $t_1$ and our preparation step \cite{Goggin2011,Robens2015,Lambert2016,Wang2017}.  Declaring $Q_1 = +1$, the LG correlator becomes
\beq
  K = \ew{Q_2} + \ew{Q_3 Q_2} - \ew{Q_3}
  \label{EQ:Kreducedfirst}
  .
\eeq 
We assume that our measurements unambiguously reveal one of $M$ different outcomes, each of which we associate with a different ``macroscopically-distinct'' state.  We  allot a $Q$-value to each via $q(n) = \pm 1$ with $1\le n \le M$ \cite{Budroni2014}.
In terms of the probabilities $P(n_i)$ of obtaining result $n_i$ at time $t_i$, the simple expectation values in $K$ read
$\ew{Q_i} = \sum_{n_i} q(n_i) P(n_i) $.  Under assumptions (A1) and (A3) [but not (A2)], adding a measurement at time $t_3$ does not affect the result at $t_2$ and so we can write $\ew{Q_2} =\sum_{n_3n_2} q(n_2) P(n_3,n_2)$, where $P(n_3,n_2)$ is the joint probability of measuring $n_2$ and $n_3$.
We then use the signalling quantifiers $ \delta(n_3)$ in \eq{EQ:firstdeltadefn} to eliminate $P(n_3)$ from $K$.  The result is
\beq
  K
  &=&
  \sum_{n_3,n_2} 
    \left[
      q(n_2) + q(n_2) q(n_3) - q(n_3) 
    \right]
    P(n_3,n_2)
    \nonumber\\
    &&
  - 
  \sum_{n_3} q(n_3) \delta(n_3)
  \label{EQ:Kwithdelta}
  .
\eeq
The first term here is what we get under the standard derivation of the LGI with NIM. The second term is new and describes the effects of the invasiveness of our measurements.
Taking a maximally-adverse position, independent maximisation of these two terms yields our modified NIM-free LGI \footnote{We
note that an alternative approach to including signalling effects in LGIs has been discussed by Kujala and Dzhafarov. See, for example, J.~V.~Kujala and E.~N.~Dzhafarov in \textit{Contextuality from Quantum Physics to Psychology} edited by E. Dzhafarov, S. Jordan, R. Zhang, and V. Cervantes (World Scientific, New Jersey, 2015), pp. 287-308.
}
\beq
  K \le 1 + \Delta
  ; \quad \text{with} \quad \Delta \equiv \sum_{n_3} |\delta(n_3)|
  \label{EQ:K3modified}
  .
\eeq
The idea is, therefore, to make measurements of both $K$ and $\Delta$ and compare them with this inequality.
It is immediately clear, however, that as long as $P(n_3,n_2)$ and $P_{3}(n_3)$ form two sets of genuine probabilities, then \eq{EQ:K3modified} can never be violated.  This holds just as well for probabilities obtained quantum-mechanically, and indeed irrespective of whether the measurements are projective or more general.

\section{Ambiguous measurements \label{SEC:ambig}}

The foregoing makes clear that we are never going to violate \eq{EQ:K3modified} with directly-measured probabilities.  The only remaining possibility is therefore to replace the measured probabilities with \textit{quasiprobabilities} in a way that a macrorealist would feel was a fair substitution.  We maintain that this can only be the case when we perform two experiments and compare them.
The first experiment proceeds as above with our detector at time $t_2$ giving unambiguously one of $1\le n\le M$ outcomes. These results are repeatable.
The second experiment analyses the same system, but with the detector at time $t_2$ being ambiguous \cite{Dressel2012}.  This detector gives one of $1 \le \alpha \le M_A$ results with the key property that repeated measurements do not necessarily lead to the same outcome.
The macrorealist will view this measurement as only revealing incomplete information about the ``real state'' of the system. 

We then look to relate the two experiments.  By following a measurement of unambiguous result $n$ with an ambiguous measurement, we obtain the conditional probabilities $c_{\alpha n}$ that state $n$ gives response $\alpha$.  Using these results and Bayes' rule, the macrorealist would be happy to  write the probability of obtaining result $\alpha$ as
\beq
  P(\alpha) = \sum_n c_{\alpha n} \mathcal{P}(n)
  \label{EQ:RcP}
  .
\eeq
We use the notation $\mathcal{P}$ to denote a probability that is not measured directly but rather inferred.
Collecting coefficients $c_{\alpha n}$ into matrix $\mathbf{c}$ and assuming $M_\mathrm{A} \ge M$ such that the ambiguous measurements give us sufficient information to reconstruct $\mathcal{P}(n)$, we write
\beq
   \mathcal{P}(n) = \sum_\alpha d_{n\alpha} P(\alpha)
  \label{EQ:PcR}
  ,
\eeq
where $d_{n\alpha}$ are elements of the left-inverse of $\mathbf{c}$, i.e. 
$\mathbf{d} \cdot \mathbf{c} = \mathbbm{1}$.  Note that we can not measure the quantities $d_{n\alpha}$ directly, as we do not know what it means to prepare in ``state $\alpha$''.

In analogy with \eq{EQ:PcR}, we next write down the inferred joint probability
\beq
  \mathcal{P}(n_3,n_2) = \sum_{\alpha} d_{n_2 \alpha}P(n_3,\alpha)
  \label{EQ:assume2}
  .
\eeq
For a macrorealist to agree that this inferred probability is same as the directly-measured one, $P(n_3,n_2)$, he or she would have to assume that the evolution of the system from state $n_2$ is the same with the ambiguous detector in place as it would be had we actually measured result $n_2$ with the unambiguous detector.  We codify this assumption as 
\begin{enumerate}
  \item[\A{2*}] \textit{Equivalently-invasive measureability} (EIM): The invasive influence of ambiguous measurements on any given macroreal state is the same as that of unambiguous ones,
\end{enumerate} 
and make it here as an alternative to NIM.

The degree of signalling due to the ambiguous measurements is quantified by
\beq
  \delta_\mathrm{A}(n_3) \equiv P(n_3) -  \sum_{\alpha} P(n_3,\alpha)
  \label{EQ:deltaA}
  .
\eeq
Inserting $\sum_n d_{n\alpha} = 1$ from the conservation of probability and using \eq{EQ:assume2}, we obtain 
\beq
  \delta_\mathrm{A}(n_3) = P(n_3) -  \sum_{n_2} \mathcal{P}(n_3,n_2)
  \label{EQ:deltaAQP}
  .
\eeq
Thus, as a consequence of (A2*), the macrorealist will expect the signalling quantifiers $\delta(n_3)$ and $\delta_\mathrm{A}(n_3)$ to be the same.  Let us quantify this by defining the ``signalling differences'' 
\beq
  D(n_3) &\equiv&  \delta(n_3) -  \delta_\mathrm{A}(n_3)
  \label{EQ:deltadiff}
  ,
\eeq
which, under EIM, obey
\beq
  D(n_3) = 0;~\forall n_3
  .
\eeq
In analogy with no-signalling in time, we dub these conditions the ``equal-signalling-in-time'' (ESIT) equalities.

To obtain an LGI that may be violated, we take \eq{EQ:Kwithdelta} and replace the measured probabilities with the inferred ones, \eq{EQ:assume2}. This yields the correlator
\beq
  K_\mathrm{A}
  &=&
  \sum_{n_3,n_2,\alpha} 
    \left[
      q(n_2) + q(n_2) q(n_3) - q(n_3) 
    \right] d_{n_2 \alpha} P(n_3,\alpha)
  \nonumber\\
  &&
  - 
  \sum_{n_3} q(n_3) \delta_\mathrm{A}(n_3)
  \label{EQ:KD}
  ,
\eeq
which involves measured quantities only. Perceiving this correlator to be equivalent to 
\eq{EQ:Kwithdelta}, the macrorealist would expect $K_\mathrm{A}$ and $\delta_\mathrm{A}(n_3)$ to be related in the same way as the original $K$ and $\delta(n_3)$.  Thus we obtain
\beq
  K_\mathrm{A} \le 1 + \Delta_\mathrm{A}
  ;\quad\quad
  \Delta_\mathrm{A} = \sum_{n_3} |\delta_\mathrm{A}(n_3)|
  \label{EQ:LGIambig}.
\eeq
This we will refer to as the ambiguously-measured LGI. The important point is that the   macrorealist would only write this inequality down if they were convinced firstly of the  existence of the states $n$ (for which we need the unambiguous measurements), and secondly that assumption (A2*) is valid.

\section{Quantum formulation\label{SEC:quantum}}

We now describe the situation quantum-mechanically.   
Let  $\rho_i$ be the system density matrix at time $t_i$, and let the time evolution from time $t_{i}$ to $t_j$ be given by $\rho_j =  \Omega_{ji}[\rho_i] = U_{ji}\rho U^\dag_{ji}$ with $U_{ji}$ the appropriate unitary operator.
Our unambiguous measurement is described with orthogonal projectors $\Pi_{n}$, $1 \le n \le M$ that obey $ \sum_{n} \Pi_{n} = \mathbbm{1}$ and $\Pi_{n}\Pi_{n'} = \delta_{n n'}\Pi_{n}$, thus ensuring the repeatability of the measurement.
In this case the signalling quantifiers are given by
\beq
  \delta ^\mathrm{QM}(n_3)
  = \sum_{n, n'\ne n} X(n_3,n,n')
  \label{EQ:deltaproj}
  .
\eeq
with 
$
  X(n_3,n,n') \equiv \trace{\Pi_{n_3}\Omega_{32}\left[\Pi_n \rho_2 \Pi_{n'}\right]}
$.

In the ambiguous case, the measurement is described by a POVM with elements that we take to be sums of projection operators
\beq
  F_\alpha = \sum_{n_2} c_{\alpha n_2} \Pi_{n_2}
  ,
\eeq
and have a clear interpretation in terms of the states $n$.  With
$
  P(\alpha) = \trace{F_\alpha \rho_2}
$ and 
$
  P(n) = \trace{\Pi_n \rho}
$, this immediately reproduces \eq{EQ:RcP}.  
The POVM elements are associated with the Kraus operators as
$F_\alpha = M_\alpha^2$ with $M_\alpha = M^\dag_\alpha =  \sum_n \sqrt{c_{\alpha n}} \, \Pi_n $, such that the joint probabilities are given by
$
  P(n_3,\alpha)
  =
  \trace{\Pi_{n_3}\Omega_{32}\left[M_\alpha \rho_2 M_\alpha\right]}
$.
This gives the ambiguous signalling quantifiers as
\beq
  \delta^\mathrm{QM}_\mathrm{A}(n_3) = \sum_{n, n'\ne n}\gamma(n,n') X(n_3,n,n')
  \label{EQ:deltaQM1}
\eeq
with
$
  \gamma(n,n')
  = 1-  \sum_\alpha \sqrt{c_{\alpha n}c_{\alpha n'}}
$.
Similarly, from \eq{EQ:KD}, the LG correlator is
\beq
  K^\mathrm{QM}_\mathrm{A} &=&
  \sum_{n_3 n_2}
  \left[
    q(n_2) + q(n_2)q(n_3) - q(n_3)
  \right]  P(n_3,n_2)
  \nonumber\\
  &&
  -\sum_{n_3}q(n_3) \delta^\mathrm{QM}_\mathrm{A}(n_3)
  \nonumber\\
  &&
  +
  \sum_{n_2,n_3}
     \left[q(n_2) + q(n_2) q(n_3) - q(n_3)\right]
  \kappa (n_3, n_2)\nonumber
  ,
\eeq
with 
\beq
  \kappa(n_2,n_3) \equiv
  \sum_{n, n'\ne n}
  \Gamma(n_2,n,n')X(n_3,n,n') 
  ,
  \label{EQ:newbit1}
\eeq
and
$
  \Gamma(n_2,n,n') 
  \equiv 
  \sum_\alpha
    d_{n_2 \alpha} \sqrt{c_{\alpha n}c_{\alpha n'}}
$.
The important result here is that, whereas the first two terms in  $K^\mathrm{QM}_\mathrm{A}$ are exactly what we get in the unambiguous case [see \eq{EQ:Kwithdelta}], a third term appears which is not directly related to the signalling quantifiers. This new term opens up the possibility of violating \eq{EQ:LGIambig}.

If we look at the signalling difference, however, we find that
\beq
  D^\mathrm{QM}(n_3) &=&  \sum_{n, n'\ne n} \left[ 1- \gamma(n,n')\right] X(n_3,n,n') 
   .
\eeq
Unless we can set these quantities to zero, the macrorealist can conclude that (A2*) does not hold, and any violation of the ambiguously-measured LGI is due to the non-comparability of the two experiments.

\section{Inverted measurements and a quantum-optics realisation\label{SEC:invmeasure}}

Is it possible to satisfy ESIT  $D(n_3)=0;~\forall n_3$ and still violate \eq{EQ:LGIambig}?  We answer this question by considering a simple measurement scheme which we call an ``inverted measurement''.
The idea is that, whereas the unambiguous detector identifies the system as being in state $n$, the inverted-measurement detector identifies it as being in any state other than $n$.
So, with three unambiguous outcomes $n\in\left\{A,B,C\right\}$, our inverted-measurement detects the three disjunctions: $A\cup B$, $B\cup C$, and $A\cup C$. From these, a macrorealist would have no qualms inferring the (quasi-)probabilities $\mathcal{P}(A) = P(A\cup B)+P(A\cup C)-P(B\cup C)$, etc. 
Such a detector has $M_\mathrm{A} = M$ and is described by the matrices
\beq
  \mathbf{c} = \frac{1}{M-1}\rb{ \mathbbm{J} - \mathbbm{1} }
  ;\quad
  \mathbf{d} = \mathbbm{J} - (M-1) \mathbbm{1} 
  \label{EQ:cmatrixinverted}
  ,
\eeq
where $\mathbbm{1}$ is the unit matrix and $\mathbbm{J}$ is a matrix of ones.

With this detector, we obtain
$
  (M-1)\delta^\mathrm{QM}_\mathrm{A}(n_3)
  = \delta^\mathrm{QM}(n_3)
$,
with $\delta^\mathrm{QM}(n_3)$ as in \eq{EQ:deltaproj}, and thus 
\beq
  D^\mathrm{QM}(n_3)=  \rb{M-2}\delta_A(n_3)
  .
\eeq
Thus, if we can find a quantum dynamics that obeys NSIT, then ESIT will be automatically satisfied.
Furthermore, with this measurement set-up, the terms responsible for LGI violations read 
\beq
  \kappa(n_3,n_2) 
   &=&
   -
   \delta^\mathrm{QM}_\mathrm{A}(n_3)
   \nonumber\\
   &&
   +\sum_{n} \left[X(n_3,n_2,n) + X(n_3,n,n_2)\right]
   ,
\eeq
which remain finite even when $\delta^\mathrm{QM}_\mathrm{A}(n_3) = 0$.
Thus this scheme offers a route to satisfy NSIT for both measurements, ESIT along with it, and still violate \eq{EQ:LGIambig}.

We now consider a three-level system as the lowest-dimensional system for which inverse measurements make sense.
We label the states $n \in \left\{A, B, C\right\}$, choose measurement assignments $q(A)=-q(B)=q(C)=1$, and initialise the system in state $\rho_1 = \op{C}{C}$.  Time evolution is governed by $U_{21}=U_{32}=U$ with 
\beq
  U &=&
  \rb{
    \begin{array}{ccc}
    1 & 0 & 0 \\
    0 & \cos \phi & \sin \phi \\
    0 & -\sin \phi & \cos \phi \\
    \end{array}
  }
  \times
  \rb{
    \begin{array}{ccc}
    \cos \chi & 0 &  \sin \chi\\
    0 & 1 & 0 \\
    -\sin \chi & 0 & \cos \chi \\
    \end{array}
  }
  \nonumber\\
  &&
  \times
  \rb{
    \begin{array}{ccc}
    \cos \theta & \sin \theta & 0 \\
    -\sin \theta & \cos \theta & 0 \\
    0 & 0 & 1 \\
    \end{array}
  }
  ,
\eeq
with parameters $\phi$, $\chi$ and $\theta$.

\begin{figure}[t]
  \begin{center}
    \includegraphics[width=0.9\columnwidth]{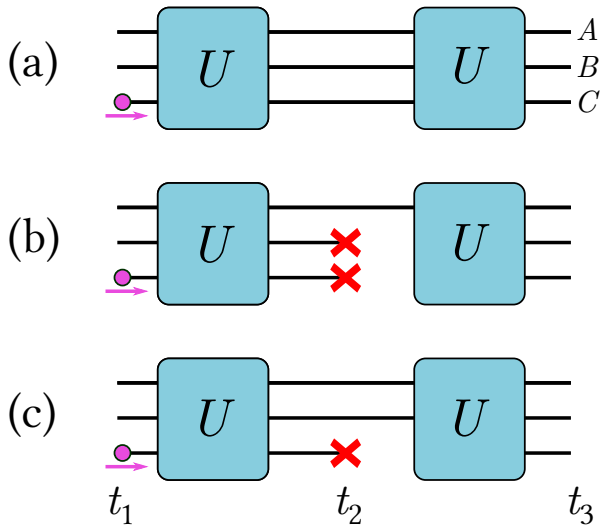}
    \caption{
      Sketch of a three-level system realised as optical channels $A$, $B$ and $C$ with non-trivial time evolution generated by the blocks labelled $U$.  We initialise by injecting a photon into channel $C$.  Three configurations are shown.
      (a) No measurement at $t_2$;
      (b) Unambiguous measurement.  With blocking elements in channels $B$ and $C$, detection of the photon at $t_3$ means that we can infer that the photon was in channel $A$ at time $t_2$.
      (c) Ambiguous measurement. With only channel $C$ blocked, detection of the photon at $t_3$ means that, from a macrorealistic point of view, the photon was in either channel $A$ or $B$ at time $t_2$.
      \label{FIG:3LS}
    }
   \end{center}
\end{figure}

A set-up like this was realised optically in Refs.\,\cite{Wang2017,Wang2017a}, and our inverted measurement scheme would have a particularly straightforward implementation in this context, see \fig{FIG:3LS}. 
The system state is encoded in one of three optical channels.  Measurements are made through a combination of photon detectors on the far right ($t_3$) and by placing a sequence of blocking elements in the optical paths at $t_2$.  
Projective measurement of the probabilities $P(n_3,A)$ are unambiguously obtained by blocking two paths at once (\fig{FIG:3LS}b) since  if we block e.g. paths $B$ and $C$, the photon must have passed through channel $A$ at $t_2$ to survive through to the detector.
In contrast, our inverted measurements are obtained by blocking just one of the three channels (\fig{FIG:3LS}c).  With a block in channel $C$, say, a detection of a photon at $t_3$ would lead a macrorealist to infer that the photon state must have been either $A$ or $B$ at time $t_2$.

\begin{figure}[t]
  \begin{center}
   \includegraphics[width=\columnwidth]{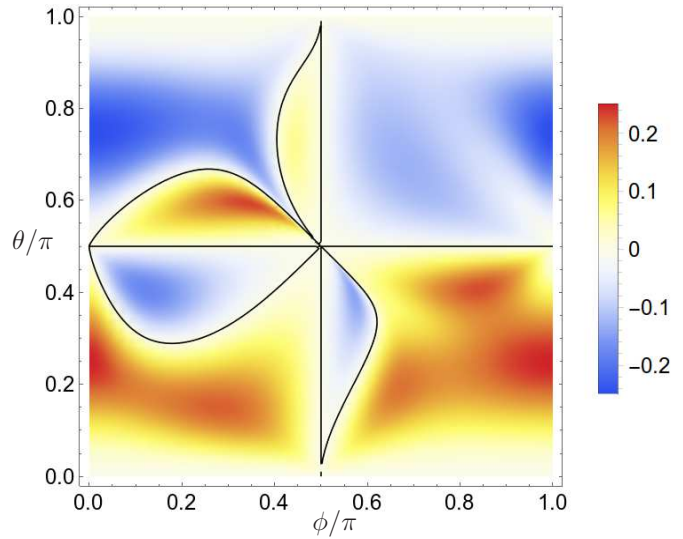}
   \caption{
    The signalling quantifier $\delta_\mathrm{A}^\mathrm{QM} (B)$ as a function of parameters $\theta$ and $\phi$ for ambiguous measurements of a three-level system. 
    Parameter $\chi$ was chosen to set $\delta_\mathrm{A}^\mathrm{QM} (A) =0$.
    The black lines indicate parameters for which $\delta_\mathrm{A}^\mathrm{QM} (B) = - \delta_\mathrm{A}^\mathrm{QM} (C) =0$ and both NSIT and ESIT are obeyed.
    \label{FIG:NSITmap}
   }
   \end{center}
\end{figure}

\begin{figure}[t]
  \begin{center}
    \includegraphics[width=\columnwidth]{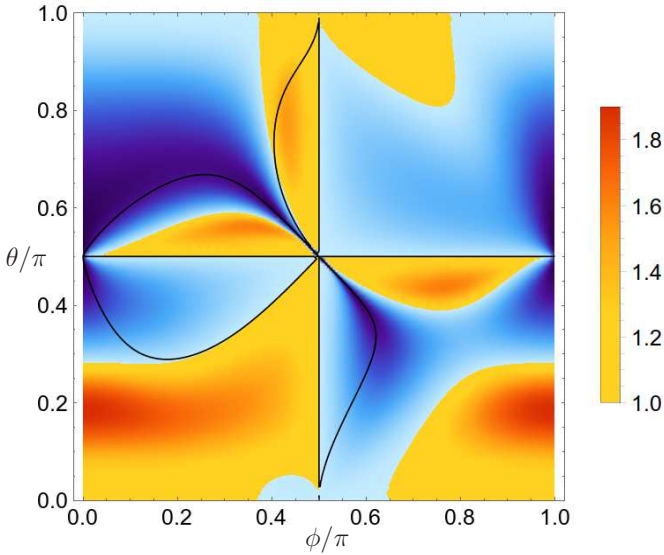}
    \caption{
      The ambiguously-measured LG correlator $K_\mathrm{A}^\mathrm{QM}$ for our three-level system as a function of angles $\theta$ and $\phi$. Red/orange colours correspond to $K_\mathrm{A}^\mathrm{QM} > 1$; blue  to $K_\mathrm{A}^\mathrm{QM} < 1$.  The black lines are the no-signalling lines from \fig{FIG:NSITmap} along which $\delta_\mathrm{A}^\mathrm{QM} (n_3) = 0;~\forall n_3$.  In the top-left quadrant we see the no-signalling line intersect a $K>1$ region, such that we have NSIT, ESIT and a violation of the LGI.
      \label{FIG:Kmap}
    }
  \end{center}
\end{figure}

We then calculate $\delta_\mathrm{A}^\mathrm{QM}$ for this system and choose $\chi$ as a function of $\phi$ and $\theta$ such that $\delta_\mathrm{A}^\mathrm{QM} (A)= 0$.  Since 
$
\sum_{n_3}\delta_\mathrm{A}^\mathrm{QM} (n_3) = 0
$, we have  $\delta_\mathrm{A}^\mathrm{QM} (B) = - \delta_\mathrm{A}^\mathrm{QM} (C)$ such that when one of these two remaining NSIT indicators is set to zero, then all three are zero.  \fig{FIG:NSITmap} shows $\delta_\mathrm{A}^\mathrm{QM} (B) $ as a function of the two angles $\theta$ and $\phi$. Marked in black are the parameters for which $\delta_\mathrm{A}^\mathrm{QM} (B) =0$.
\fig{FIG:Kmap} shows the corresponding LGI correlator $K_\mathrm{A}^\mathrm{QM}$,  which takes values up to a maximum of $K^\mathrm{QM}_\mathrm{A}=1.9$.  Overlaid on this figure are the no-signalling lines from \fig{FIG:NSITmap}.   In many places, these lines coincide with the LG correlator taking the value $K_\mathrm{A}^\mathrm{QM}=1$.  However, there are also several regions where this is not the case, and in particular, in the top left corner of this figure we see a no-signalling line intersect a region with $K_\mathrm{A}^\mathrm{QM}>1$. For these parameter values, then, we have NSIT, ESIT and a violation of the LGI.

\fig{FIG:Kfollow} shows two cuts through \fig{FIG:Kmap}. \fig{FIG:Kfollow}a reveals the maximum value of $K_\mathrm{A}^\mathrm{QM}$ when signalling is zero to be $K_\mathrm{A}^\mathrm{QM}=1.464$.
\fig{FIG:Kfollow}b shows a straight cut through\fig{FIG:Kmap} for fixed $\theta$. On this plot we also show the quantity $1+\Delta_\mathrm{A}^\mathrm{QM}$, which represents the modified upper bound for $K_\mathrm{A}^\mathrm{QM}$.  Only around the points where $\Delta^\mathrm{QM}_\mathrm{A}$ is close to zero do we obtain LGI violations.

\section{Weak measurements \label{SEC:weak}}

As example of LGI violations with weak measurements, let us consider a detector ($M_\mathrm{A} = M$) described by
\beq
  \mathbf{c} = \frac{1-\epsilon}{M} \mathbbm{J} + \epsilon \mathbbm{1} 
  ;\quad\quad
  \mathbf{d} = \frac{1}{\epsilon} \mathbbm{1} + \frac{\epsilon-1}{\epsilon M}\mathbbm{J}
  .
\eeq
Each detector response is biased towards a certain (unambiguous) outcome,  but in the limit $\epsilon \to 0$ this bias disappears and the measurement becomes weak.
To leading order in $\epsilon$, we obtain
$\gamma(n, n') \approx \frac{M}{4} \epsilon^2$ and $\Gamma(n_2, n, n') \approx \frac{1}{2} \rb{ \delta_{n,n_2} + \delta_{n',n_2}}$.  Thus, with this detector in the $\epsilon \to 0$ limit, the ambiguous NSIT quantities become
\beq
  \lim_{\epsilon \to 0} \delta_\mathrm{A}^\mathrm{QM}(n_3) = 0
  ,
\eeq
and there is no signalling for the weak measurement.
Meanwhile, for terms responsible for violation of the ambiguous LGI, we obtain
\beq
  \lim_{\epsilon \to 0} 
  \kappa(n_2,n_3) 
  = 
  \frac{1}{2} \sum_{n\ne n_2} \left[ X(n_3,n_2,n)+ X(n_3,n,n_2) \right]
  \nonumber
   ,
\eeq
which will be non-zero provided there are coherences between basis states at time $t_2$.  Indeed, with these results we can rewrite the LG correlator as
\beq
  \lim_{\epsilon \to 0} K_\mathrm{A}^{QM} 
  &=& 
  \trace{
    \left[
      \hat{Q}_2
      + \textstyle{\frac{1}{2}}\left\{\hat{Q}_2,\hat{Q}_3\right\}
      -\hat{Q}_3
    \right]
  \rho_1}
  \label{EQ:Kweak}
\eeq
where $\hat{Q}_n = \hat{Q}(t_n) = U^\dag_{n1} \rb{\sum_m q(m) \Pi_m} U_{n1}$ is the measured operator in the Heisenberg picture at time $t_n$ and $\left\{\cdot,\cdot\right\}$ denotes the anticommutator.  We thus arrive at the weakly-measured form of the LGI as discussed in e.g. \citer{Halliwell2016}.  From Fritz \cite{Fritz2010}, we know that the maximum quantum-mechanical value of this quantity is identical to that obtained in the projective case in the L\"uders limit \cite{Budroni2013,Budroni2014}, i.e. when the number of projectors is exactly two. Thus, we conclude that, in the weak-measurement case,
\beq
  \lim_{\epsilon \to 0} K_\mathrm{A}^{QM} \le 1 + \Delta_\textnormal{{\scriptsize L\"uders}}^\mathrm{QM} \le \frac{3}{2},
  \label{EQ:weakLueders}
\eeq
where $\Delta_\textnormal{{\scriptsize L\"uders}}^\mathrm{QM}$ is the no-signalling quantity that would be obtained under a projective L{\"u}ders measurements.  Since this will generally be non-zero, no-signalling violations of the weakly-measured LGI are possible.

\begin{figure}[t]
  \begin{center}
   \includegraphics[width=\columnwidth]{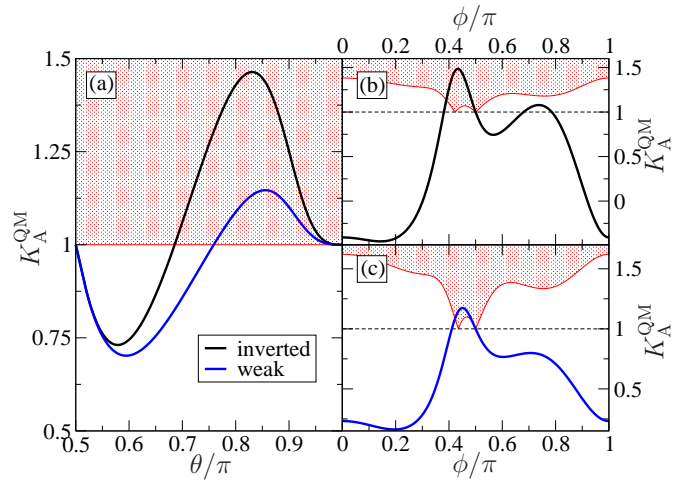}
    \caption{
      Two cuts through \fig{FIG:Kmap} for the inverted-measurement case plus corresponding results for the weak measurements of \secref{SEC:weak}.
      \textbf{(a)} The correlator $K_\mathrm{A}^\mathrm{QM}$ as a function of $\theta$ along the no-signalling line that goes through the orange region in \fig{FIG:Kmap}.  Since $\Delta_\mathrm{A} = 0$ along this line, the LGI reverts to $K_\mathrm{A} \le 1$ and violations of \eq{EQ:LGIambig} occur within the indicated red region.  Black line: inverted measurement; blue line: weak measurement. 
      \textbf{(b)} The correlator $K_\mathrm{A}^\mathrm{QM}$ along the straight-line cut in  \fig{FIG:Kmap} from $\phi=0$ to $\phi = \pi$ with $\theta = 0.831 \pi$. The red region shows the righthand side of \eq{EQ:LGIambig} and, since generally we have signalling here, this quantity is greater than one. Indeed, only near the maximum of $K_\mathrm{A}^\mathrm{QM}$ does $1+ \Delta_\mathrm{A}^\mathrm{QM}$ drop significantly such that we obtain a violation of \eq{EQ:LGIambig}.
      NB: the maximum of the $K_\mathrm{A}^\mathrm{QM}$ curve here is slightly displaced from the no-signalling point and thus has a value slightly higher than the no-signalling maximum ($1.482$ vs. $1.464$).
      \textbf{(c)} The same as panel (b) but for the weakly-measured case  with  $\theta =  0.856 \pi$.  Again the maximum is slightly offset from no-signalling maximum ($1.173$ vs. $1.147$).
      \label{FIG:Kfollow}
    }
  \end{center}
\end{figure}

This, however, indicates a problem when the number of outcomes for our unambiguous measurements is actually $M=2$, because then the quantity $\Delta_\textnormal{{\scriptsize L\"uders}}^\mathrm{QM}$ in \eq{EQ:weakLueders} is exactly the same $\Delta^\mathrm{QM}$ for the unambiguous measurements. Thus violations of the ambiguously-measured LGI imply violations of the  unambiguous NSIT equalities.  In the $M=2$ case, therefore, we have $\lim_{\epsilon\to0} D^\mathrm{QM}(n_3) = \delta^\mathrm{QM}(n_3) \ne 0$ for at least some $n_3$, from which the realist would conclude that (A2*) is invalid.

Away from this $M=2$ case, however, this argument does not apply, and we may obtain $\Delta^\mathrm{QM} =0$ whilst $\Delta_\textnormal{{\scriptsize L\"uders}}^\mathrm{QM} >0$, since they are different quantities.
To show that is the case, we return to the three-level system of the last section for which $M=3$.  For this model, we already know that  $\Delta^\mathrm{QM} = \Delta^\mathrm{QM}_\mathrm{A} = 0$ along the lines shown in \fig{FIG:NSITmap}. A plot of the weakly-measured LG correlator of \eq{EQ:Kweak} [not shown] then looks very similar to \fig{FIG:Kmap} (but with less-pronounced maxima) and again shows a region of LGI violation intersected by the no-signalling line. 
\fig{FIG:Kfollow}a shows the value of $K_\mathrm{A}^\mathrm{QM}$ along this line, from which we obtain a maximum violation of $K_\mathrm{A}^\mathrm{QM}=1.147$ (and thus  $\Delta_\textnormal{{\scriptsize L\"uders}}^\mathrm{QM} = 0.147$).
\fig{FIG:Kfollow}c also shows a cut for fixed $\theta$ through the maximum.

\section{Discussion \label{SEC:disc}}

In \secref{SEC:noNIM} we saw that in attempting to derive an LGI without the NIM assumption (A2), we ensure that it can never be violated.  This is not surprising because a realistic description of nature by itself is not inconsistent with the probabilities of quantum mechanics \cite{Leggett2008}. 
To obtain an inequality that we could violate, we considered two measurements, one unambiguous and one ambiguous \footnote{
An interesting extension would be to consider more than two experiments with differing degrees and/or types of ambiguity}. By comparing the two, and replacing a probability in an inequality derived for one experiment with a (quasi-) probability inferred from  the other, we obtained our ``ambiguously-measured LGI'' which, as we have seen in our examples, can be violated by quantum theory. 
To justify this switch, however, we had to introduce the EIM assumption (A2*), in which the invasivity of the measurements in each of the two experiments was taken to be equivalent.  Making the role of this assumption explicit in LGI tests with ambiguous or weak measurements is one of the main results of this work.

Whilst the EIM assumption may not seem particularly plausible in the abstract \footnote{EIM does, however, follow automatically from NIM, so it can be no less plausible}, for certain detectors, it might be.
In particular, consider measurements that are realised with a set of individual detectors, each of which only interacts with the system when it is in just one of the macro-real states (as in ideal negative measurements). In this case, then, the unambiguous measurement would be implemented by using one of these detectors at a time, whereas the ambiguous measurement would involve using more than one, deployed in such a way that any knowledge of which particular detector had fired was lost.  Since in this case, the components of the ambiguous measurement are ``just the sum'' of those from the unambiguous one, one could reasonably expect the influence on the system of the two measurements to be the same.
But, more than consideration of any specific realisation, just as NIM leads to the NSIT conditions, so the EIM assumption leads to the ESIT conditions, and these can be experimentally tested.  And while successful NSIT/ESIT tests can not exclude NIM/EIM, they would at least provide the macrorealist with some level of empirical confidence that the experiment was functioning in conformity with these principles.

In this paper we have discussed the concrete example of a three-level system, under both inverted and weak measurements, and found parameter regimes where it can satisfy both NSIT and ESIT equalities whilst violating an LGI.
In the weak measurement case, it is important to note that whilst the NSIT equality is guaranteed to hold for the weak measurement itself, this is not necessarily the case for the unambiguous part of the experiment and thus, generally, ESIT would not hold in these experiments.  Only under certain model-specific circumstances, and only when the system dimension is greater than two, can both NSIT and ESIT be fulfilled.
As we show in appendix\,\ref{SEC:3box}, the three-box problem from the experiment of George et al. \cite{George2013} can be understood within the framework discussed here and shows the required NSIT/ESIT properties.

Faced, then, with the violation of an ambiguously-measured LGI, together with the satisfaction of ESIT, what would a macrorealist conclude?  
Certainly, this would would give more cause for thought than having measured an LGI violation in a single experiment, as there a measurement of the NSIT equalities would be enough to dismiss the measurements as signalling.
With the ambiguous prescription and ESIT, the macrorealist would be faced with either giving up the combination of A1+A3, or finding an explanation for how two different measurements can somehow conspire to give exactly the same degree of signalling and yet somehow influence the system in very different ways.  This problem is compounded in the case where ESIT is satisfied through NSIT being satisfied for both experiments, as in this case both experiments are individually non-signalling. 
With this, then, we arrive at a situation similar to that presented by Wilde and Mizel \cite{Wilde2012} (with a recent realisation \cite{Huffman2016}).  Whilst the measurement procedures in their work were very different to those considered here (involving different bases), the problem created for the  macrorealist is similar --- in order to maintain a macrorealistic description of the system,  the macrorealist is left with having to explain away a collusion between two sets of measurements.  
When properly executed, then, ambiguously-measured LGIs provide a further way with we might narrow the ``clumsiness loophole''.


\begin{acknowledgments}
We are grateful to George Knee, Jonathan Halliwell, KunKun Wang and Peng Xue for helpful discussions.
\end{acknowledgments}

\appendix
\section{Quantum three-box problem \label{SEC:3box}}

The quantum three-box problem \cite{George2013,Maroney2012} can be cast in the language used here.  With the three states labelled as in \secref{SEC:invmeasure}  Alice's measurement at time $t_3$ is characterised by two projectors
$\Pi^{(3)}_{-1} = \op{C}{C}$ and $\Pi^{(3)}_{+1} = \mathbbm{1} - \op{C}{C}$,
which we have labelled with the respective $q=\pm 1$ assignments and a superscript to distinguish them from the previous projectors.
Bob's measurement at time $t_2$ can be characterised by a four-element POVM with 
\beq
  \mathbf{c} = \frac{1}{2}
  \rb{
    \begin{array}{ccc}
      1 & 0 & 0\\
      0 & 1 & 1\\
      0 & 1 & 0\\
      1 & 0 & 1
    \end{array}
  }
  ,
\eeq
and the assignments $q(A)=q(B)=-q(C)=+1$.  For the time evolution operators we take
\beq
 U_{21}=
 \frac{1}{\sqrt{6}}
 \rb{
  \begin{array}{ccc}
    2 & 0 & \sqrt{2} \\
    -1 & \sqrt{3} & \sqrt{2} \\
    -1 &- \sqrt{3} & \sqrt{2}
  \end{array}
  };
  \\
 U_{32}=
 \frac{1}{\sqrt{6}}
 \rb{
  \begin{array}{ccc}
    1 & 1 & 2 \\
    \sqrt{3} & -\sqrt{3} & 0\\
    \sqrt{2} & \sqrt{2} & - \sqrt{2}
  \end{array}
 }
 .
\eeq
Finally, we need to consider a different, but essentially equivalent \cite{Emary2014}, version of the LGI:
\beq
  K' = \ew{Q_2}_{21} + \ew{Q_3 Q_2}_{321} - \ew{Q_3}_{31} \ge -1
  .
\eeq
Calculating with the above formalism gives
$\delta^{QM}_A(n_3) = \delta^{QM}(n_3) = 0$ such that Bob's measurements are non-signalling and ESIT is satisfied, along with  
$
 {K'}^\mathrm{QM}_\mathrm{A} = -13/9
$,
such that the ambiguous LGI is violated.  This is in agreement with \citer{George2013}.

\input{LGI_ambiguous_final.bbl}

\end{document}

%% file: LGI_ambiguous_final.bbl
%